\documentclass[12pt]{article}

\usepackage{amssymb,amsmath}

\usepackage{graphicx}
\usepackage{subfigure}

\newcommand{\be}{\begin{equation}}
\newcommand{\ee}{\end{equation}}
\newcommand{\Dlt}{\Delta}
\newcommand{\dlt}{\delta}

\newcommand{\br}{{\bf r}}
\newcommand{\bn}{{\bf n}}
\newcommand{\bbe}{{\bf e}}
\newcommand{\bB}{{\bf B}}
\newcommand{\bS}{{\bf S}}
\newcommand{\bt}{\beta}

\newcommand{\al}{\alpha}
\newcommand{\ra}{\rightarrow}

\newcommand{\gm}{\gamma}
\newcommand{\om}{\omega}
\newcommand{\Om}{\Omega}
\newcommand{\Gm}{\Gamma}

\newcommand{\lbd}{\lambda}

\newcommand{\lgl}{\langle}
\newcommand{\rgl}{\rangle}

\begin{document}

\begin{center}

{\Large{\bf Possibility of superradiance by magnetic nanoclusters } \\ [5mm]

V.I. Yukalov$^{1,*}$ and E.P. Yukalova$^{2}$} \\ [3mm]

{\it 
$^1$Bogolubov Laboratory of Theoretical Physics, \\
Joint Institute for Nuclear Research, Dubna 141980, Russia \\ [3mm]

$^2$Laboratory of Information Technologies, \\
Joint Institute for Nuclear Research, Dubna 141980, Russia} 

\end{center}

\vskip 2cm

\begin{abstract}

The possibility of realizing spin superradiance by an assembly of magnetic
nanoclusters is analyzed. The known obstacles for realizing such a coherent 
radiation by magnetic nanoclusters are their large magnetic anisotropy, 
strong dephasing dipole interactions, and an essential nonuniformity of 
their sizes. In order to give a persuasive conclusion, a microscopic theory 
is developed, providing an accurate description of nanocluster spin dynamics. 
It is shown that, despite the obstacles, it is feasible to organize such a 
setup that magnetic nanoclusters would produce strong superradiant emission. 
\end{abstract}

\vskip 2cm


\vskip 3cm

{\bf $^*$Corresponding author e-mail}: yukalov@theor.jinr.ru

\newpage

\section{Introduction}

Coherent radiation by resonant atoms is well known and widely studied [1-5].
Different regimes of coherent optical radiation can be realized, including
{\it superradiance} that is a self-organized coherent radiation. Quantum 
dots, that are called artificial atoms, can also produce superradiance [6,7], 
as well as several other resonant finite-level systems of different physical 
nature [8-12]. Spin assemblies, as such, can radiate coherently only being 
driven by an external magnetic field, but they cannot produce superradiance, 
that is a self-organized process, because of the dephasing role of their 
dipole spin interactions [13,14]. However, superradiance can be achieved in 
spin systems if they are coupled to a resonator formed by a resonant electric 
circuit [15]. Pure superradiance by nuclear spins was, first, observed in 
Dubna experiments [16-19] and later confirmed by other experimental groups 
[20-22]. Pulsing coherent radiation by nuclear spins, under permanent pumping, 
has also been observed [23-25]. The full theory of nuclear spin superradiance 
has been developed [26-31], being in good agreement with experiment and 
numerical simulations [32-34]. 

It is necessary to stress the principal difference between the role of particle
interactions in the process of atomic and spin superradiance. Resonant atoms, 
by exchanging photons through the common radiation field, develop effective
dipole interactions that correlate atoms, leading to the radiation coherence. 
So, atomic superradiance does not require the compulsory presence of a resonator. 
Contrary to this, dipole interactions between spins dephase spin motion, thus, 
resulting in their decoherence. As a result, spin superradiance cannot be 
realized without a resonator, whose feedback field correlates the spin motion 
and leads to superradiance. Additional information on atomic superradiance can 
be found in recent works [35-37] and the literature cited there. Superradiance by 
Bose-condensed atoms has also been studied [38]. The basic difference of atomic 
superradiance from spin superradiance is thoroughly explained in Refs. [13,14].

Nuclear spin systems are usually formed of nuclei with low spins of order $1/2$. 
There also exist molecular magnets, composed of identical molecules with magnetic 
moments of order $10$ [39,40]. It has been shown theoretically [41-46] that spin 
superradiance can be realized by such molecular magnets.

It has also been mentioned [46] that another natural candidate for trying to 
realize spin superradiance would be an ensemble of magnetic nanoparticles.
General properties of the latter have been reviewed in Refs. [47,48]. 
Magnetic particles below a critical size cannot support more than one 
domain and thus are described as single-domain coherent clusters. For 
typical material parameters, the critical diameter is between $1-100$ nm, 
hence such clusters are necessarily of nanosizes. Most of the behavior 
of these clusters can be interpreted by treating them as objects where 
all magnetic moments of their internal atoms are rigidly aligned, forming 
a single giant spin that can reach the values as large as of order 
$S\sim 10^8$. If an ensemble of many such giant spins could radiate 
coherently, this would result in very high radiation intensity.

However, the realization of superradiance by a system of magnetic 
nanoparticles confronts the obstacles that seem to hinder the possibility 
of this realization. These obstacles are as follows. First, having large 
spins, these nanoparticles also have strong dipole interactions that 
dephase coherent spin motion. Second, magnetic nanoparticles possess 
rather large magnetic anisotropy, both longitudinal and transverse, 
which makes their effective transition frequencies time dependent, 
complicating by this resonance tuning [41-44]. This is contrary to proton 
spins in polarized targets, having no magnetic anisotropy [49]. Third, and 
seemingly the most unpleasant, it is practically impossible to prepare many 
nanoclusters of the same size. Their sizes usually vary in a large diapason, 
which implies a wide spin variation. This could lead to a large nonuniform 
broadening suppressing coherent radiation. The broad size distribution of 
nanoclusters makes them principally different from magnetic molecules, 
all of which can be made identical to each other and forming a molecular 
magnet with an ideal crystalline structure.    

In the present paper, the problem is studied whether it would be 
admissible to organize such a setup that would allow for the realization 
of spin superradiance by a system of magnetic nanoparticles. Since this 
is a rather intricate problem, the consideration, in order to be persuasive, 
will be based on a microscopic theory, but not on phenomenological equations.

\section{Evolution of spin variables}

Let us consider $N$ nanoparticles in a sample of volume $V$. The microscopic 
spin Hamiltonian can be presented as the sum 
\be
\label{1}
 \hat H = \sum_j \hat H_j \; + \; 
\frac{1}{2} \sum_{i\neq j} \hat H_{ij} \;  ,
\ee
the first term corresponding to the collection of single clusters, enumerated 
by the index $j = 1,2,\ldots, N$, and the second term describing their dipole 
interactions. 

Each magnetic particle of a size below the critical one behaves as a 
single-domain cluster with uniform magnetization that experiences coherent 
rotation. Magnetic anisotropy of such a nanoparticle originates from 
magnetocrystalline anisotropy, from shape anisotropy that is the anisotropy 
of the magnetostatic energy of the sample induced by its nonspherical shape, 
and from surface anisotropy entering the bulk term because of the assumption 
of uniform magnetization. Well below the Curie temperature, the average 
magnetization of a cluster can be assumed to be of constant length, so that 
the magnetization vector rotates as a whole, which is termed the coherent 
rotation. These features of small single-domain particles are the basis of 
the Stoner-Wohlfarth model [50-52]. The quantum variant of this model can 
be represented by the single-cluster Hamiltonian    
\be
\label{2}
 \hat H_j = - \mu_0 \bB \cdot \bS_j - D_j \left ( S_j^z \right )^2 +
D_{2j} \left ( S_j^x \right )^2 \;  ,
\ee
where $\mu_0=-2\mu_B =-\hbar\gamma_e$, with $\mu_B$ being the Bohr magneton 
and $\gamma_e$, the electron gyromagnetic ratio, and where $\bB$ is an external 
magnetic field, $D_j$ and $D_{2j}$ are the longitudinal and transverse anisotropy 
parameters for a $j$-th cluster, and $S_j^\alpha$ are spin operators. Sometimes, 
instead of $S_j^x$ one writes $S_j^y$, which is, evidently, the same.

In addition to the second-order anisotropy terms, one considers the 
fourth-order,
$$
D_4 \left [  
\left  ( S_j^x \right )^2 \left ( S_j^y \right )^2  +
\left ( S_j^y \right )^2  \left ( S_j^z \right )^2 + 
\left ( S_j^z \right )^2 \left ( S_j^x \right )^2
\right ]  \; ,
$$  
and sixth-order,
$$
D_6 \left ( S_j^x \right )^2 \left ( S_j^y \right )^2 
\left ( S_j^z \right )^2 \;  ,
$$
anisotropy terms. However, these terms are usually much smaller than the 
second-order ones and, to a good approximation, can be neglected.

The dipole interactions are described by the Hamiltonian 
\be
\label{3}
 \hat H_{ij} = \sum_{\al\bt} D_{ij}^{\al\bt} S_i^\al S_j^\bt \;  ,
\ee
with the dipole tensor given by the equations
$$
D_{ij}^{\al\bt} = \frac{\mu_0^2}{r_{ij}^3} \; \left ( \dlt_{\al\bt} -
3 n_{ij}^\al n_{ij}^\bt \right ) \; , \qquad
r_{ij} \equiv | \br_{ij} | \; , \qquad 
\bn_{ij} \equiv \frac{\br_{ij}}{r_{ij}} \; , \qquad 
\br_{ij} \equiv \br_i - \br_j \;  .
$$

As has been stressed and explained in detail in Refs. [13,14,41-44], a spin
system, because of the presence of dephasing dipole interactions, cannot 
produce superradiance without being coupled to a resonant electric circuit.
It is, therefore, necessary to place the considered sample into a coil of
the circuit. Then the magnetic field 
\be
\label{4}
 \bB = B_0 \bbe_z + H \bbe_x  
\ee
includes, in addition to a constant external magnetic field $B_0$, the 
feedback field $H$ of a resonator. This feedback field is described by the 
Kirchhoff equation 
\be
\label{5}
 \frac{dH}{dt} + 2\gm H + \om^2 \int_0^t H(t') \; dt' = - 4 \pi \eta \; 
\frac{dm_x}{dt} \;  ,
\ee
in which $\gamma$ is the circuit damping, $\omega$ is the circuit natural  
frequency, $\eta = V/V_{coil}$ is a filling factor, $V_{coil}$ is the coil
volume, and 
$$
m_x = \frac{\mu_0}{V} \; \sum_j \; \lgl \; S_j^x \; \rgl \;   ,
$$
with the angle brackets defining a statistical averaging.

For spin systems composed of proton or electron spins, it would be sufficient 
to tune the resonator natural frequency close to the spin Zeeman frequency 
\be
\label{6}
 \om_0 \equiv - \; \frac{\mu_0}{\hbar} \; B_0 = 
2 \; \frac{\mu_B}{\hbar}\; B_0 = \gm_e B_0 \;  .
\ee
But for the nanoclusters, the situation is more complicated. 

Deriving the evolution equations for the nanocluster spins, we will need the 
notation for the fluctuating spin parts
$$
\xi_i^0 \equiv \frac{1}{\hbar} \; \sum_{j(\neq i)} \left ( a_{ij} S_j^z +
c_{ij}^* S_j^- + c_{ij} S_j^+ \right ) \;  , 
$$
\be
\label{7}
\xi_i \equiv 
\frac{i}{\hbar} \; \sum_{j(\neq i)} \left ( 2 c_{ij} S_j^z \; - 
\; \frac{1}{2} \; a_{ij} S_j^- + 2 b_{ij} S_j^+ \right ) \; ,
\ee
with the relation between the spin components
$$
 S_j^\pm = S_j^x \pm i S_j^y \; , \qquad 
S_j^x = \frac{1}{2} \left ( S_j^+ + S_j^- \right ) \; , \qquad
S_j^y = - \; \frac{i}{2} \left ( S_j^+ - S_j^- \right ) \;  , 
$$
and where the interaction parameters are 
$$
 a_{ij} \equiv D_{ij}^{zz} \; , \qquad b_{ij} \equiv \frac{1}{4} \left ( 
D_{ij}^{xx} - D_{ij}^{yy} - 2i D_{ij}^{xy} \right ) \; ,\qquad 
c_{ij} \equiv \frac{1}{2} \left ( D_{ij}^{xx} - i D_{ij}^{yz} \right ) \; .   
$$
Also, let us define the effective force acting on a $j$-th spin,
\be
\label{8}
 f_j \equiv -\; \frac{i}{\hbar} \; \mu_0 H + \xi_j \;  .
\ee
 
Using the commutation relations
$$
[ S_i^x ,\; S_j^y ] = i \dlt_{ij} S_j^z \; , \qquad
 [ S_i^y ,\; S_j^z ] = i \dlt_{ij} S_j^x \; , \qquad
[ S_i^z ,\; S_j^x ] = i \dlt_{ij} S_j^y \; ,
$$
$$
[ S_i^+ ,\; S_j^- ] = 2 \dlt_{ij} S_j^z \; , \qquad
[ S_i^z ,\; S_j^+ ] = \dlt_{ij} S_j^+ \; , \qquad
[ S_i^z ,\; S_j^- ] = - \dlt_{ij} S_j^- 
$$
gives the following commutators involving the anisotropy terms:
$$
\left [ S_i^- ,\; ( S_j^z )^2 \right ] = S_j^- S_j^z + S_j^z S_j^- \; , 
$$
$$
\left [ S_i^- ,\; ( S_j^x )^2 \right ] = -( S_j^x S_j^z + S_j^z S_j^x) \; , 
$$
$$
\left [ S_i^z ,\; ( S_j^x )^2 \right ] = i ( S_j^x S_j^y + S_j^y S_j^x) \; . 
$$
As a result, the Heisenberg equations of motion yield the equations for the 
spin variables, transverse,
\be
\label{9}
 \frac{dS_j^-}{dt} = - i ( \om_0 +\xi_j^0 ) S_j^- + f_j S_j^z +
\frac{i}{\hbar} \; D_j\left ( S_j^- S_j^z + S_j^z S_j^- \right ) +
\frac{i}{\hbar} \; D_{2j} \left ( S_j^x S_j^z + S_j^z S_j^x \right ) \;  ,
\ee
and the longitudinal one,
\be
\label{10}
 \frac{dS_j^z}{dt} = - \;
\frac{1}{2} \; \left ( f_j^+ S_j^- + S_j^+ f_j \right ) +
\frac{1}{\hbar}\; D_{2j} \left ( S_j^x S_j^y + S_j^y S_j^x \right ) \; .
\ee

Equation (9) shows that, if the anisotropy terms would be smaller than the 
term containing $\omega_0$, then all spins, though being different, 
nevertheless would rotate with the frequencies close to $\omega_0$. Thence 
the coherent spin motion could be realized.

\section{Averaged equations of motion}

Different nanoparticles can have different spins $S_j$, hence different 
vectors ${\bf S}_j$ for which ${\bf S}_j^2 = S_j (S_j + 1)$. But we are 
interested in the coherent properties of the system, that is, in the 
behavior of the ensemble of nanoparticles as a whole. It is, therefore,
reasonable to consider the quantities averaged over all system components.
We define the average nanocluster spin
\be
\label{11}
 S \equiv \frac{1}{N} \sum_{j=1}^N \; S_j  
\ee
and the average anisotropy constants
\be
\label{12}
 D \equiv \frac{1}{N} \sum_{j=1}^N D_j \; , \qquad 
D_2 \equiv \frac{1}{N} \sum_{j=1}^N D_{2j} \;  .
\ee

Accomplishing the statistical averaging of Eqs. (9) and (10), we keep in 
mind that the radiation wavelength, $\lambda$, corresponding to the Zeeman 
frequency $\omega_0$, is larger than the intercluster distance $a$. Under
this assumption $\lambda > a$, the mean-field approximation is admissible.
Thus, for spins on different sites, we set
\be
\label{13}
 \lgl \; S_i^\al S_j^\bt \; \rgl =   \lgl \; S_i^\al  \; \rgl
 \lgl \; S_j^\bt \; \rgl  \qquad ( i \neq j) \; ,
\ee
where $\alpha$ and $\beta$ are arbitrary Cartesian indices. For the spins, 
of different components but on the same site, the correct approximation [41-44] 
reads as
\be
\label{14}
  \lgl \; S_j^\al S_j^\bt + S_j^\bt S_j^\al\; \rgl = 
\left ( 2 - \; \frac{1}{S_j} \right )  
\lgl \; S_j^\al \; \rgl  \lgl \; S_j^\bt \; \rgl  \qquad ( \al \neq \bt) \; .
\ee
This formula interpolates between the spin $S_j = 1/2$, when Eq. (14) becomes 
the identity $0=0$, since the Pauli matrices anticommute, and large spins, for
which the mean-field approximation is asymptotically exact, as $S_j\ra\infty$.

The averaged spin variables of interest are the {\it transition function}
\be
\label{15}
u \; \equiv \; \frac{1}{SN} \sum_{j=1}^N \; \lgl \; S_j^-  \; \rgl \;  ,
\ee
the {\it coherence intensity}
\be
\label{16}
 w \; \equiv \;
\frac{1}{S^2N(N-1)} \; \sum_{i\neq j}^N \; \lgl \; S_j^+ S_j^- \; \rgl \; ,
\ee
and the {\it spin polarization}
\be
\label{17}
 s \; \equiv \; \frac{1}{SN} \sum_{j=1}^N \; \lgl \; S_j^z  \; \rgl \;  .
\ee
   
To retain the attenuation effects, coming from the dipole spin interactions, one 
includes in the equations the dephasing rate that, taking into account saturation 
effects [41-44], reads as
\be
\label{18}
 \Gm_2 = \gm_2( 1 - s^2) \qquad (\gm_2 = \rho \hbar \gm_e^2 S ) \;  ,
\ee
with $\rho \equiv N/V$ being the density of nanoparticles. These nanoclusters 
are immersed into a matrix, whose molecules interact with spins. Thence, it is 
also necessary to include the related relaxation rate $\gamma_1$. In the case 
of the existence of a pumping polarization mechanism, the relaxation rate is 
$\Gamma_1 = \gamma_1 + \gamma_1^*$, where $\gamma_1^*$ is a pumping rate.

Other averaged quantities of importance are the average fluctuating fields
\be
\label{19}
\xi_0 \; \equiv \; \frac{1}{N} \sum_{j=1}^N \; \lgl \; \xi_j^0  \; \rgl \; , 
\qquad
\xi \; \equiv \; \frac{1}{N} \sum_{j=1}^N \; \lgl \; \xi_j  \; \rgl \; ,
\ee
and the average effective force 
\be
\label{20}
 f \equiv \frac{1}{N} \sum_{j=1}^N \; \lgl \; f_j  \; \rgl =
- \; \frac{i}{\hbar} \; \mu_0 H + \xi \;  .
\ee

Let us introduce the {\it longitudinal anisotropy frequency}
\be
\label{21}
\om_D \equiv ( 2S - 1) \; \frac{D}{\hbar}
\ee
and the {\it transverse anisotropy frequency}
\be
\label{22}
\om_2 \equiv ( 2S - 1) \; \frac{D_2}{\hbar} \;   .
\ee
And let us define the {\it effective anisotropy frequency}
\be
\label{23}
\om_A \equiv \om_D + \frac{1}{2}\; \om_2
\ee
and the {\it effective rotation frequency}
\be
\label{24}
 \Om \equiv \om_0 - \om_A s \;  .
\ee
We shall also need the notation
\be
\label{25}
 F \equiv f + \frac{i}{2} \; \om_2 u^* = -\;
\frac{i}{\hbar} \; \mu_0 H + \xi + \frac{i}{2} \; \om_2 u^* \;  .
\ee
  
Averaging Eqs. (9) and (10), we get the equations for the average 
variables (15) to (17):
$$
\frac{du}{dt} = - i ( \Om + \xi_0 - i \Gm_2 ) u + Fs \; , \qquad
\frac{dw}{dt} = - 2 \Gm_2 w + \left ( u^* F + F^* u \right ) s \; ,
$$
\be
\label{26}
 \frac{ds}{dt} = - \; \frac{1}{2}\; \left  ( u^* F + F^* u \right ) - 
\Gm_1( s - \zeta) \; ,
\ee
where $\zeta$ is the equilibrium polarization of a single cluster. 

The Kirchhoff equation (5) can be represented [26-29] as the integral 
{\it feedback equation}
\be
\label{27}
 H = - 4\pi \eta \int_0^t G(t-t') \dot{m}_x(t') \; dt' \;  ,
\ee
with the transfer function
$$
G(t) = \left [ \cos(\widetilde\om t) \; - \;
\frac{\gm}{\widetilde\om} \; \sin (\widetilde\om t) \right ] \; 
e^{-\gm t} \; ,
$$
in which $\widetilde{\omega} \equiv \sqrt{\omega^2 - \gamma^2}$, and with the 
source term
$$
\dot{m}_x \equiv \frac{1}{2} \; \rho \mu_0 S \; \frac{d}{dt} (u^* + u) \;  .
$$
As is seen from these equations, the characteristic parameter defining the 
strength of the feedback field is the {\it feedback rate}
\be
\label{28}
 \gm_0 \equiv \frac{\pi}{\hbar} \; \eta \rho \mu_0^2 S \;  .
\ee

To analyze further Eqs. (26), we employ the scale separation approach [26-30,53]
that is a variant of the averaging techniques [54,55]. The fast oscillating 
fields (19) are treated as stochastic variables, with the stochastic averaging
$$
\lgl \lgl \; \xi_0(t) \; \rgl \rgl = 
\lgl \lgl \; \xi(t) \; \rgl \rgl = 0 \; , \qquad
\lgl \lgl \; \xi_0(t) \xi_0(t') \; \rgl \rgl = 
2\gm_3 \dlt(t-t') \; ,
$$
\be
\label{29}
 \lgl \lgl \; \xi_0(t) \xi(t') \; \rgl \rgl = 
\lgl \lgl \; \xi(t) \xi(t') \; \rgl \rgl = 0 \;  , \qquad
\lgl \lgl \; \xi^*(t) \xi(t') \; \rgl \rgl = 2\gm_3 \dlt(t-t') \; ,
\ee
in which $\gamma_3$ is the dynamic broadening of the order of $\gamma_2$. 

Microscopic evolution equations (9) and (10), as well as the averaged equations
(26), show that different spins can rotate with the approximately same frequency, 
close to $\omega_0$, only when their dipole interactions and magnetic anisotropies 
are sufficiently small, such that they do not disturb much the spin motion. And the
feedback action of the resonator is effective, if the circuit natural frequency 
$\omega$ is tuned close to the Zeeman frequency $\omega_0$. The latter implies
the resonance condition
\be
\label{30}
 \left | \; \frac{\Dlt}{\om} \; \right | \ll 1 \qquad 
( \Dlt \equiv \om - \om_0 ) \;  .
\ee
While the condition that the effective rotation frequency $\Omega$ be close to 
the Zeeman frequency $\omega_0$ requires that the anisotropy be small, such that
\be
\label{31}
 \left | \; \frac{\om_D}{\om} \; \right | \ll 1 \qquad  
\left | \; \frac{\om_2}{\om} \; \right | \ll 1 \; .
\ee
Also, all attenuation rates are assumed to be smaller than the natural 
frequency $\omega$, that is, 
$$
\frac{\gm}{\om} \ll 1 \; , \qquad  \frac{\gm_0}{\om} \ll 1 \; , \qquad 
\frac{\gm_1 }{\om} \ll 1 \; , 
$$
\be
\label{32}
\frac{\gm_2}{\om} \ll 1 \; , \qquad \frac{\gm_3 }{\om} \ll 1 \;  .
\ee
   
Under inequalities (32), the feedback equation (27) can be solved by iterating 
its right-hand side with $u(t) = u(0) \exp (-i \Omega t)$, which yields
\be
\label{33}
 \mu_0 H = i \hbar ( \al u - \al^* u^* ) \;  ,
\ee
where the coupling function $\alpha$, for a sharp resonance, such that 
$|\Delta/\gamma| < 1$, reads as
\be
\label{34}
 \al = g \gm_2 \; \frac{\Om}{\om_0} \left ( 1 - e^{-\gm t} \right ) \;  ,
\ee
with the dimensionless {\it coupling parameter}
\be
\label{35}
 g \equiv \frac{\gm \gm_0 \om_0}{\gm_2(\gm^2+\Dlt^2)} \;  .
\ee
In what follows, we shall also need the effective coupling function
\be
\label{36}
 \widetilde\al \equiv \al + \frac{i}{2} \; \om_2 \;  .
\ee
 
Substituting the feedback field (33) into Eqs. (26) yields the equations
\be
\label{37}
 \frac{du}{dt} = - i ( \Om + \xi_0 ) u - ( \Gm_2 - \al s) u +
\xi s - s ( \widetilde\al u )^* \; ,
\ee
\be
\label{38}
\frac{dw}{dt} = -2  ( \Gm_2 - \al s) w + (u^*\xi + \xi^* u ) s - 
2s {\rm Re}( \widetilde\al u^2 ) \; ,
\ee
\be
\label{39}
 \frac{ds}{dt} = - \al w \; - \; \frac{1}{2} (u^*\xi + \xi^* u ) -
\Gm_1 ( s - \zeta ) + {\rm Re}( \widetilde\al u^2 ) \;  .
\ee
Complimenting these by the initial conditions $u_0 = u(0)$, $w_0 = w(0)$, 
$s_0 = s(0)$, without the loss of generality, one can set $u_0 = u_0^*$ 
to be real.

Equations (37) to (39) show that the variable $u$ can be treated as fast,
as compared to the slow variables $w$ and $s$. Following the scale separation 
approach [26-30,53], we solve Eq. (37) for the fast variable $u$, keeping there
the slow variables $w$ and $s$ as quasi-integrals of motion, which gives
$$
 u = u_0 \exp \left \{ - (i\Om + \Gm_2 - \al s) t -
i \int_0^t \xi_0(t') \; dt' \right \} \; +
$$
$$
+ \; s \int_0^t \xi(t') \exp \left \{ - (i \Om + \Gm_2 - \al s) ( t - t') -
i \int_{t'}^t \xi_0(t'')\; dt'' \right \} \; dt' \;  .
$$
Substituting this solution into Eqs. (38) and (39), we average the right-hand 
sides of the latter over time and over the stochastic variables (19), again 
keeping the slow variables fixed. In this way, we come to the equations for 
the guiding centers:   
\be
\label{40}
 \frac{dw}{dt} = -2  ( \Gm_2 - \al s) w + 2\gm_3 s^2 \; , \qquad
\frac{ds}{dt} = - \al w - \gm_3 s - \Gm_1 (s -\zeta) \;  .
\ee

\section{Coherent radiation by nanoclusters}

We shall study the derived equations (40), setting there $\zeta = -1$,
assuming the exact resonance, with $\Delta = 0$, and measuring time in
units of $1/\gamma_2$. Then Eqs. (40) reduce to
\be
\label{41}
 \frac{dw}{dt} = -2  ( 1 - s^2 - \al s) w + 2\gm_3 s^2 \; , \qquad
 \frac{ds}{dt} = - \al w - \gm_3 s - \gm_1 (s + 1 ) \; .
\ee
The coupling function (34) can be written as
\be
\label{42}
 \al = g ( 1 - As) \left ( 1 - e^{-\gm t} \right ) \;  ,
\ee
where the {\it anisotropy parameter} is 
\be
\label{43}
 A \equiv \frac{\om_A}{\om_0} \;  .
\ee
   
Solving these equations, we can calculate the radiation intensity
$$
 I_S(t) = \frac{2\mu_0^2}{3c^3} \left | \; 
\sum_i \; \lgl \; \ddot{S}_j(t) \; \rgl \; \right |^2 \; ,   
$$
due to spin radiation, and the time-averaged intensity
$$
\overline{I}_S(t) =  \frac{2\mu_0^2}{3c^3} \; N^2 S^2 \Om^4 w(t) \;  .
$$
It is convenient to define the dimensionless radiation intensity
\be
\label{44}
  I(t) \equiv 
\frac{3c^3 \overline{I}_S(t)}{2\mu_0^2\om_0^4 S^2N^2} \; ,
\ee
which reduces to the form
\be
\label{45}
I(t) = [\;  1 - As(t) \; ]^4 w(t) \;  .
\ee
 
The typical behavior of solutions is shown in Figs. 1-4. The 
attenuation parameters are taken as $\gm=10$, $\gm_1=10^{-3}$, $\gm_3=1$, 
and the coupling parameter is $g=100$. Different initial conditions are 
investigated and the anisotropy parameter (43) is varied. Note that, 
according to definitions (15) to (17), it should be: $w + s^2 \leq 1$. 
Therefore, we take the initial conditions that satisfy the relation 
$w_0 + s_0^2 = 1$. It is possible to separate four qualitatively different 
regimes.

{\it Pure spin superradiance}. Figure 1 presents the case of pure spin 
superradiance, when there is no initial coherence, $w_0 = 0$, and the system 
is fully polarized, $s_0 = 1$. Increasing the anisotropy shifts the curves  
to the right. It is interesting that the maximum of the radiation intensity 
increases with $A$, which is due to the greater contribution of the first 
factor in Eq. (45).  

{\it Spin subradiance}. If there is no initial coherence imposed, $w_0 = 0$, 
and the system is not polarized at the initial time, $s_0 = -1$, then the 
values of the solutions are much smaller and the maximal radiation intensity 
is an order smaller than in the previous case of the polarized sample. This 
is shown in Fig. 2. A temporary radiation pulse happens because of the spin 
motion caused by spin interactions with each other and with the feedback field, 
though this radiation is very weak. 

{\it Spin induction}. When one imposes on the system strong initial coherence, 
$w_0 = 1$, with no polarization, $s_0 = 0$, then the regime of spin induction 
is realized, as is illustrated in Fig. 3. The variation of the anisotropy does 
not influence much the behavior of $w$ and $s$, but essentially increases the 
radiation intensity, with increasing $A$, and shifts the radiation maximum to 
nonzero time. While in the absence of anisotropy, the maximum of radiation 
intensity is at the initial time $t=0$. 

{\it Triggered spin superradiance}. The intermediate situation, shown in 
Fig. 4, corresponds to triggered spin superradiance, when there exists an 
essential initial polarization, $s_0 = 0.9$, and the spin motion is triggered 
by an initial coherent pulse, $w_0 = 0.19$. The maximum of the radiation 
intensity again becomes larger, when increasing $A$.

These results demonstrate that an ensemble of nanoclusters can produce coherent 
radiation. When coherence is imposed on the sample at the initial moment of 
time, the regimes of spin induction or triggered superradiance are produced. 
But, what is more important, it is feasible to realize such a setup, when the 
regime of pure spin superradiance can be achieved, as is illustrated in Fig. 1.

\section{Typical properties of nanoclusters}

Now we need to understand whether there really exist nanoclusters for which 
superradiance could be realized. As has been stressed above, nanoparticles
possess rather strong magnetic anisotropy that is usually characterized by
the anisotropy parameters, whose relation with the parameters, discussed in
Sec. 2, is as follows:
$$
 K_1 = \frac{DS^2}{V_1} \; , \qquad  
K_2 = \frac{D_2S^2}{V_1} \; , \qquad 
K_4 = \frac{D_4S^4}{V_1} \; , 
$$
where $V_1$ is the volume of a single nanoparticle. Respectively, 
$$
D = \frac{K_1V_1}{S^2} \; , \qquad  D_2 = \frac{K_2V_1}{S^2} \; , 
\qquad D_4 = \frac{K_4V_1}{S^4} \;   .
$$
The density of a single cluster is $\rho_1 = N_1/V_1$, with $N_1$ being the 
number of atoms composing the cluster. The total spin of a cluster is 
proportional to the number of atoms, $S \sim N_1$. Therefore the reduced
anisotropy parameters are   
$$
 D \sim \frac{K_1}{\rho_1 N_1} \; , \qquad 
D_2 \sim \frac{K_2}{\rho_1 N_1} \; , \qquad
D_4 \sim \frac{K_4}{\rho_1 N_1^3} \;  .
$$
There exists a great variety of clusters with different properties [47,48].
Below, we shall discuss some of them in more detail.

{\bf Co clusters} [56,57]. The coherence radius $R_{coh}$, below which coherent 
rotation of magnetization takes place, is $R_{coh} \approx 10$ nm. The radii
below which particles are single-domain are about three times larger. But of 
the major importance for us is the coherence radius. The usual radius of a 
single cluster is $R \approx 1-2$ nm, containing the number of cobalt atoms 
$N_1 \approx 1000-1500$, the cluster volume being $V_1 \approx 2 \times 10^{-20}$ 
cm$^3$. The anisotropy parameters 
$$
K_1 = 0.22 \; {\rm MJ/m}^3 = 2.2 \times 10^6   {\rm erg/cm}^3 \; , 
$$
$$
K_2 = 0.09 \; {\rm MJ/m}^3 = 0.9 \times 10^6   {\rm erg/cm}^3 \; ,
$$
$$
K_4 = -0.01 \; {\rm MJ/m}^3 = -0.1 \times 10^6   {\rm erg/cm}^3 \;
$$
are caused by shape anisotropy, magnetocrystalline anisotropy, and surface 
anisotropy. The magnetic moment is well defined below the blocking temperature
$T_B \approx 14$ K.

{\bf Fe clusters} [57]. The coherence radius is $R_{coh} \approx 7.5$ nm. The 
number of iron atoms in a cluster is $N_1 \approx 1000$. The typical size of 
a single cluster is similar to that of a Co cluster. The anisotropy parameters are
$$
K_1 = 0.32 \; {\rm MJ/m}^3 = 3.2 \times 10^6   {\rm erg/cm}^3 \; , 
\qquad
K_4 = 0.05 \; {\rm MJ/m}^3 = 0.5 \times 10^6   {\rm erg/cm}^3 \; ,
$$
with $K_2 \approx 0$.

{\bf Ni clusters} [58]. The average radii of clusters are $R \approx 3-4$ nm.
The blocking temperature is $T_B \approx 20-40$ K. Magnetic anisotropy is 
characterized by $K_1 = 1.5 \times 10^6$ erg/cm$^3$.

{\bf CoAg clusters} [57]. The characteristic radii are close to those of the 
above clusters. Anisotropy parameters are
$$
K_1 = 0.02 \; {\rm MJ/m}^3 = 2 \times 10^5   {\rm erg/cm}^3 \; , 
\qquad
K_2 = 0.006 \; {\rm MJ/m}^3 = 0.6 \times 10^5   {\rm erg/cm}^3 \; .
$$
Similar anisotropy parameters characterize CoNb an CoPt clusters [57], with
the blocking temperature of order 100 K. 

{\bf CoFe$_2$O$_4$ clusters} [59]. Typical cluster radii can vary in the range
of $1-10$ nm. Depending on the radius of a cluster, its anisotropy parameters
can be as follows:
$$
K_1 = 1.9 \times 10^7 {\rm erg/cm}^3 \qquad (R = 2.85 \;{\rm nm} )\; , 
$$
$$
K_1 = 1.4 \times 10^7 {\rm erg/cm}^3 \qquad (R = 3.50\; {\rm nm} )\; ,
$$
$$
K_1 = 2.5 \times 10^6 {\rm erg/cm}^3 \qquad (R = 6.35\; {\rm nm} )\; ,
$$
The blocking temperature is about $200-300$ K.

{\bf Fe$_3$0$_4$ clusters} [60]. This material is called magnetite. 
Clusters are of the radius $R \approx 2-3$ nm. The magnetic anisotropy 
is
$$
 K_1 = 0.36 \; {\rm MJ/m}^3 = 3.6 \times 10^6   {\rm erg/cm}^3 \;  .
$$
The blocking temperature is $T_B \approx 45$ K.

There exist many other clusters [47,48,61] with a variety of properties. 
For estimates, we shall take the values typical of Co, Fe, and Ni 
clusters. The coherence radius for these clusters is $R_{coh} = 10-100$ nm. 
Respectively, the number of atoms in a cluster of $10$ nm is $N_1\sim 10^5$, 
while in a cluster of $100$ nm, it is $N_1 \sim 10^8$. The standardly 
prepared clusters have the radii $R \sim 1-3$ nm. The corresponding cluster 
volume is $V_1\sim 10^{-20}$ cm$^3$. The number of atoms in a cluster is 
$N_1\sim 10^3$. The related density inside a cluster is $\rho_1\sim 10^{23}$ 
cm $^{-3}$. The cluster spin is $S \sim 10^3-10^5$, which gives the magnetic 
moment of the order of $10^3-10^5 \mu_B$. The blocking temperature, when 
thermally activated reversals are suppressed is $T_B \sim 10-100$ K. The 
magnetic anisotropy parameters are $K_1 \sim K_2 \sim 10^6$ erg/cm$^3$, 
while $K_4$ is an order smaller. The reduced anisotropy parameters are 
$D \sim D_2 \sim 10^{-20}$ erg, while $D_4 \sim 10^{-27}$ erg. The ensemble 
of clusters is placed in a matrix, so that the density of clusters is 
$\rho \sim 10^{20}$ cm$^{-3}$.

Using these values gives for the anisotropy frequencies (21), (22), 
and (23) $\omega_D \sim \omega_2 \sim \omega_A \sim 10^{10}$ 1/s. This 
corresponds to the anisotropy field $B_D \equiv \omega_D / \gamma_e$. 
With $\gm_e\sim 10^7$ 1/G s, $\mu_0\sim 10^{-20}$ erg/G, and G$^2$=erg/cm$^3$, 
we have $B_D\sim 10^3$ G. The feedback rate (28) is $\gamma_0 \sim 10^{10}$ 
1/s. The dephasing rate, defined in Eq. (18), is $\gamma_2 \sim 10^{10}$ 1/s. 
As we see, the anisotropy frequencies are of the order of the attenuation 
rates $\gm_0$ and $\gm_2$, and the anisotropy parameter (43) is $A\sim 0.1$. 
Therefore the resonance conditions (32) can already be satisfied for the 
Zeeman frequency (6) about $\om_0\sim 10^{11}$ 1/s. For this purpose, it 
is sufficient to have an external magnetic field $B_0 \sim 1$ T. The 
wavelength, corresponding to this $\om_0$ is $\lbd\sim 1$ cm. The mean 
intercluster distance is $a=1/\rho^{1/3}\sim 10^{-7}$ cm. Hence $\lbd\gg a$, 
and the employed mean-field approximation is well justified. 

The maximal radiation intensity, for the accepted parameters can reach 
very high values $I_S \sim 10^{19}$ erg/s, that is, $I_S \sim 10^{12}$ W. 
The superradiance pulse is very short, being $t_p \sim 1/g \gamma_2$. Here 
it is $t_p \sim 10^{-12}$ s. But it could be made much shorter, since the 
coupling parameter (35) is of the order $g \sim \omega_0 / \gamma$, hence, 
diminishing the resonator damping $\gamma$ the pulse duration could be 
made $\ll 10^{-12}$ s.

In conclusion, we have demonstrated that, despite the presence of anisotropy, 
strong dipole dephasing interactions, and nonuniform sizes, it is feasible 
to organize such a setup, when an ensemble of nanoclusters would produce a 
short superradiant pulse of high intensity. The developed theory is based 
on a microscopic Hamiltonian and the analysis involves realistic parameters 
typical of many magnetic nanoclusters. Fast polarization reversal of magnetic
nanoclusters can find many applications, e.g., in magnetic recording [62,63],
for magnetization-reversal measurements with micro-SQUID techniques [64], 
in quantum information processing, quantum computing, and for creating quantum 
artificial intelligence [65].

\vskip 5mm
{\it Acknowledgement}. Financial support from the Russian Foundation for 
Basic Research is appreciated.

\newpage

\begin{center}
{\Large{\bf Figure Captions}}
\end{center}

\vskip 3cm

{\bf Fig. 1}. Regime of pure spin superradiance. Numerical solutions of Eqs. 
(40) for the coherence intensity $w(t)$, spin polarization $s(t)$, and the 
radiation intensity $I(t)$ as functions of time $t$ (in units of $1/\gm_2$). 
The attenuation parameters (in units of $\gm_2$) are $\gm=10$, $\gm_3=1$, and 
$\gm_1 = 10^{-3}$. The coupling parameter is $g = 100$. The initial conditions 
are $w_0 = 0$ and $s_0 = 1$.

\vskip 1cm
{\bf Fig. 2}. Regime of spin subradiance. Coherence intensity $w(t)$, spin 
polarization $s(t)$, and radiation intensity $I(t)$ for the same system 
parameters, as in Fig. 1, but for the initial conditions $w_0=0$, $s_0=-1$, 
corresponding to nonpolarized clusters.

\vskip 1cm
{\bf Fig. 3}. Regime of spin induction. Coherence intensity $w(t)$, spin 
polarization $s(t)$, and radiation intensity $I(t)$ for the same parameters, 
as in Fig. 1, but for the initial conditions $w_0 = 1$, $s_0 = 0$.

\vskip 1cm
{\bf Fig. 4}. Regime of triggered spin superradiance. Coherence intensity 
$w(t)$, spin polarization $s(t)$, and radiation intensity $I(t)$ for the 
same parameters, as in Fig. 1, but for the initial conditions $w_0 = 0.19$, 
$s_0 = 0.9$.

\newpage

\begin{figure}[ht]
\vspace{9pt}
\centerline{
\hbox{ \includegraphics[width=8cm]{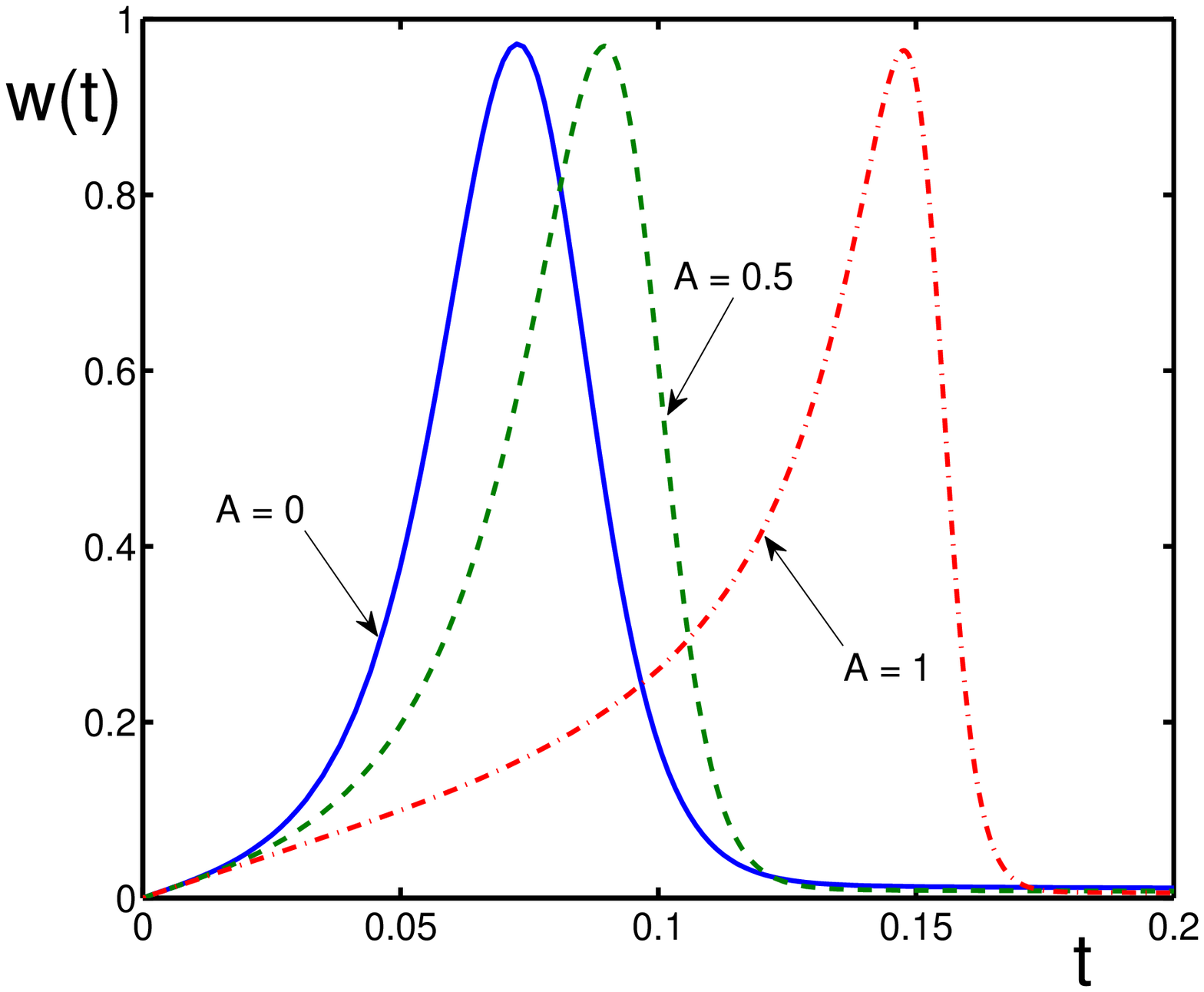} \hspace{1cm}
\includegraphics[width=8cm]{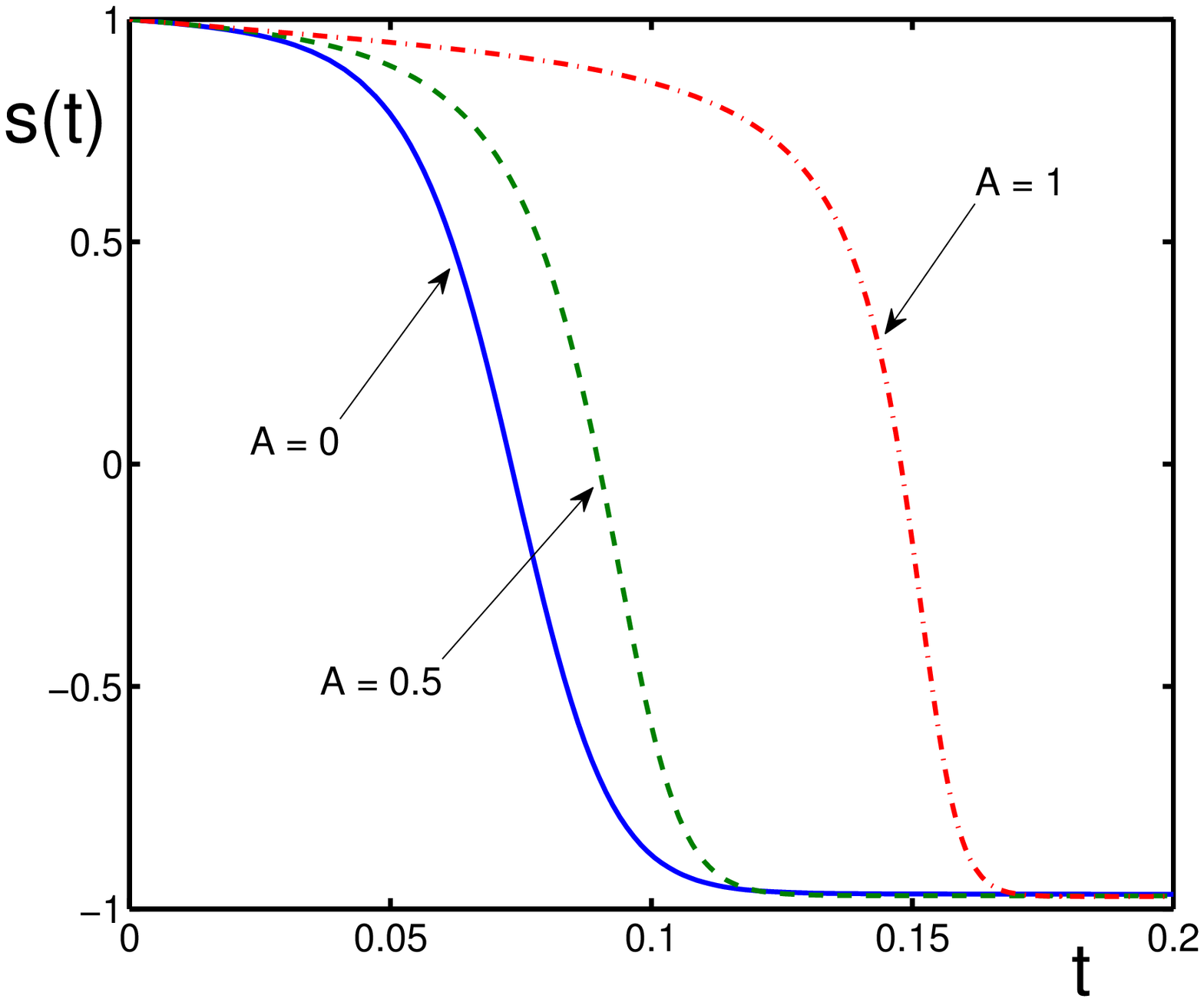} } }
\vspace{9pt}
\centerline{
\hbox{ \includegraphics[width=8cm]{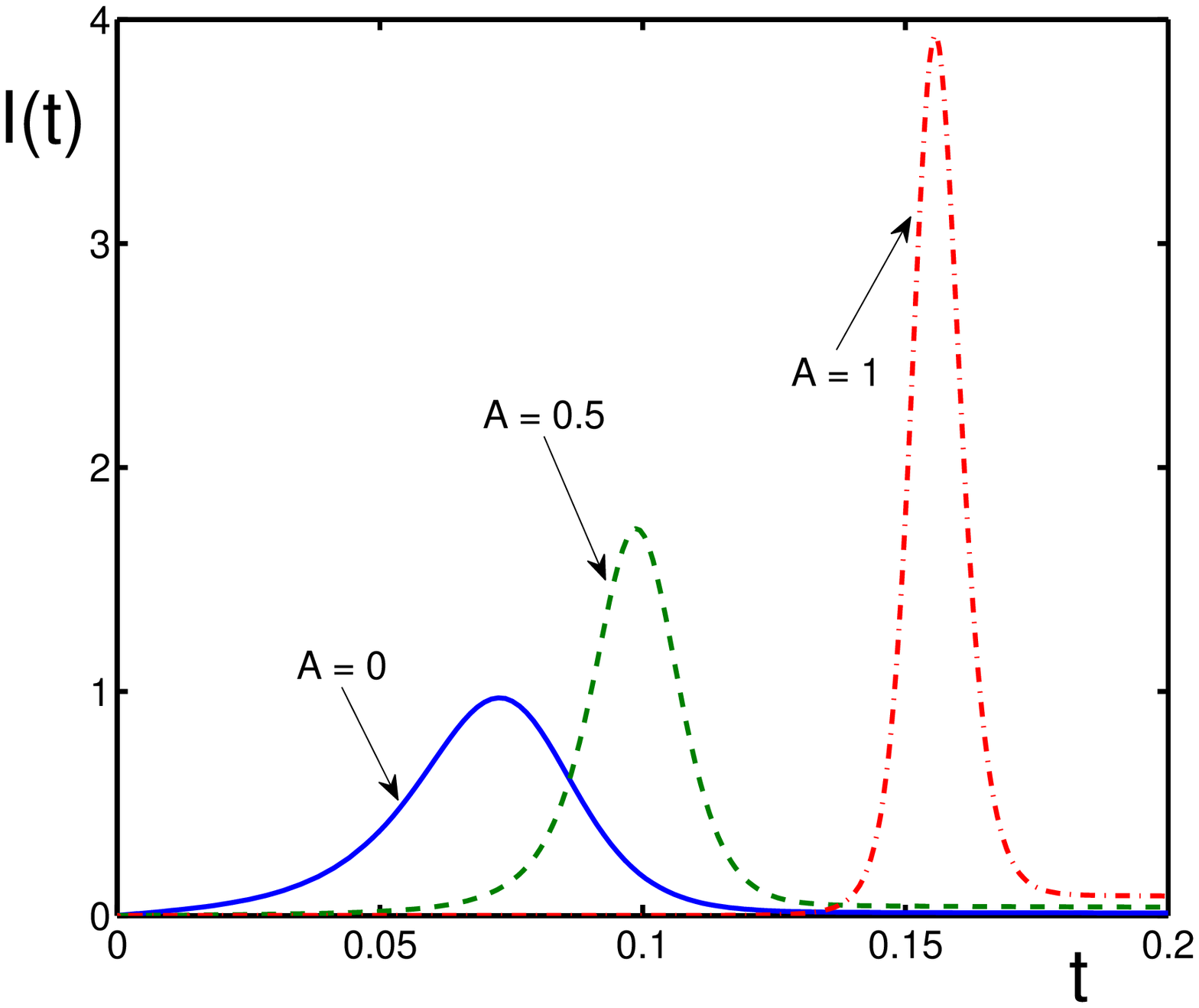} } }
\caption{Regime of pure spin superradiance. Numerical solutions of Eqs. 
(40) for the coherence intensity $w(t)$, spin polarization $s(t)$, and 
the radiation intensity $I(t)$ as functions of time $t$ (in units of 
$1/\gm_2$). The attenuation parameters (in units of $\gm_2$) are $\gm=10$, 
$\gm_3=1$, and $\gm_1 = 10^{-3}$. The coupling parameter is $g = 100$. 
The initial conditions are $w_0 = 0$ and $s_0 = 1$.}
\label{fig:Fig.1}
\end{figure}

\newpage

\begin{figure}[ht]
\vspace{9pt}
\centerline{
\hbox{ \includegraphics[width=8cm]{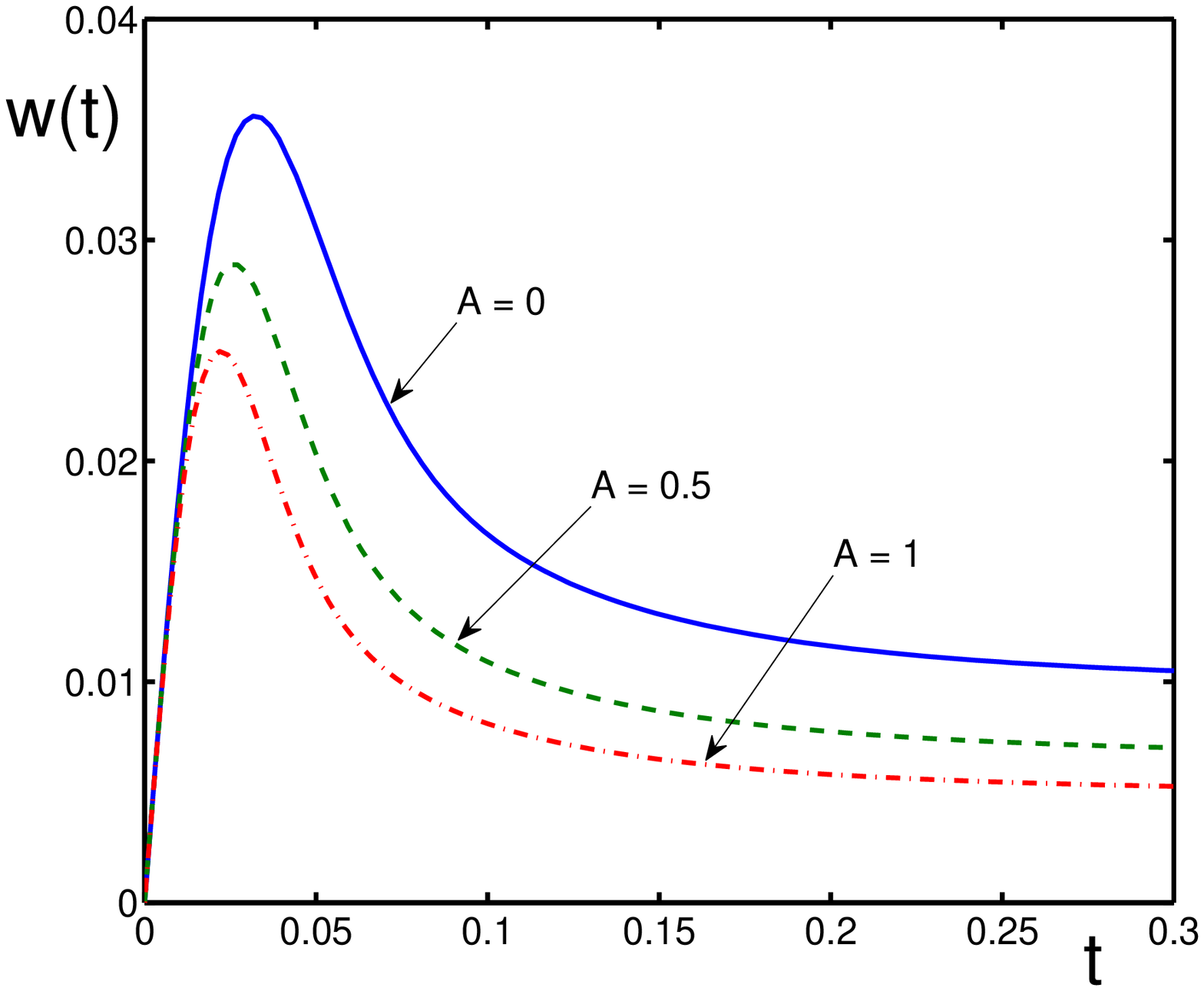} \hspace{1cm}
\includegraphics[width=8cm]{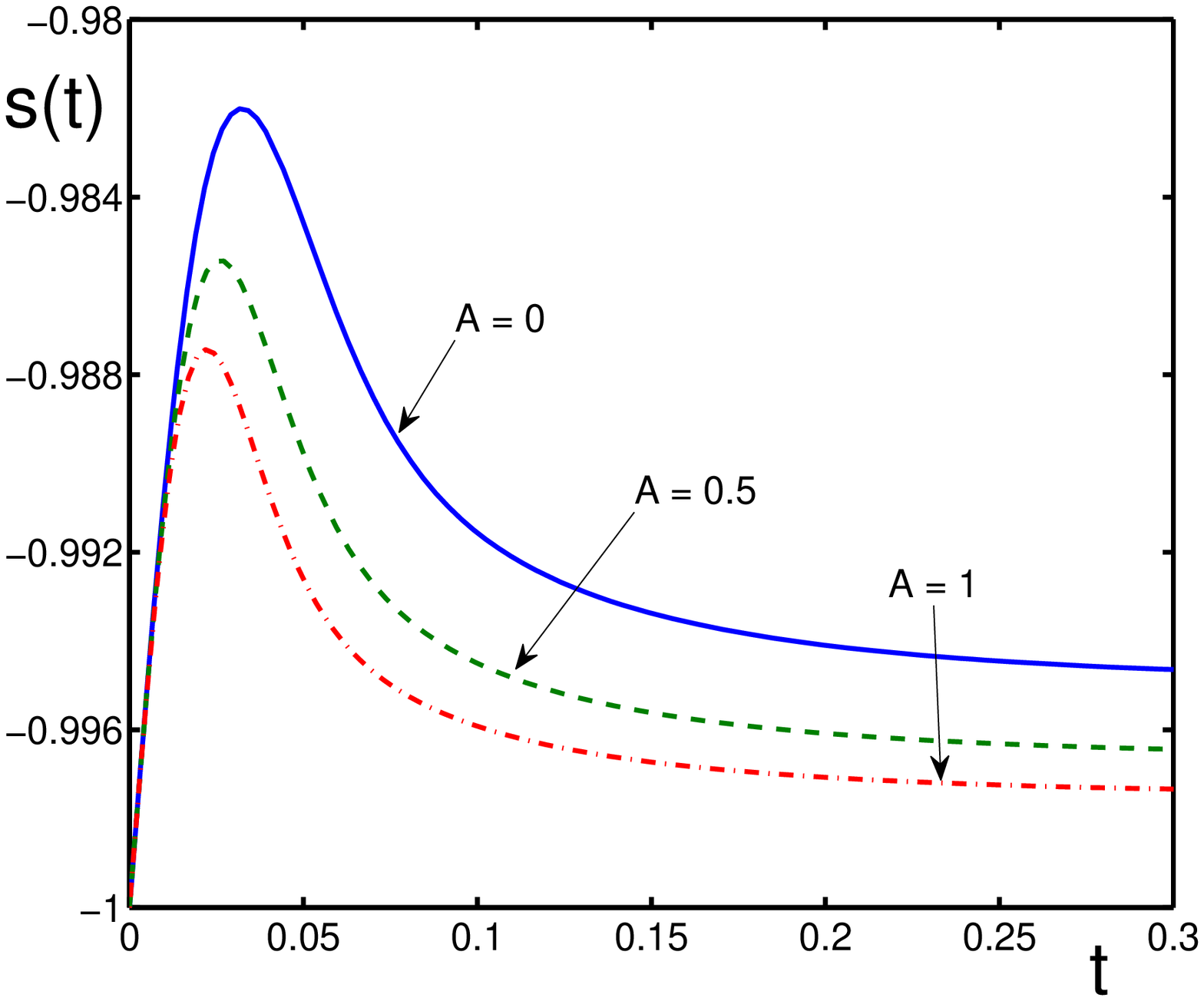} } }
\vspace{9pt}
\centerline{
\hbox{ \includegraphics[width=8cm]{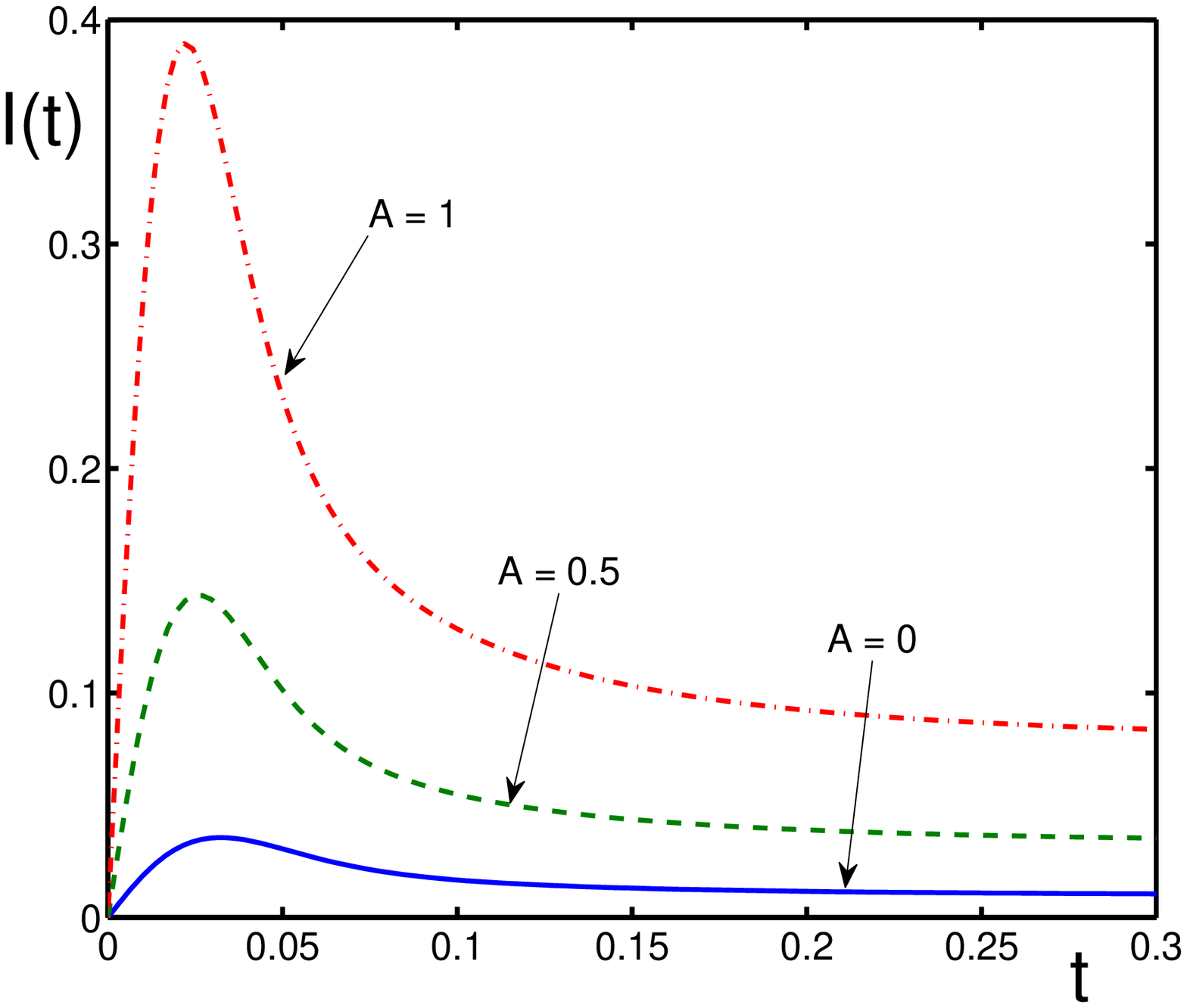} } }
\caption{ Regime of spin subradiance. Coherence intensity $w(t)$, spin 
polarization $s(t)$, and radiation intensity $I(t)$ for the same system 
parameters, as in Fig. 1, but for the initial conditions $w_0=0$, 
$s_0=-1$, corresponding to nonpolarized clusters.}
\label{fig:Fig.2}
\end{figure}

\newpage

\begin{figure}[ht]
\vspace{9pt}
\centerline{
\hbox{ \includegraphics[width=8cm]{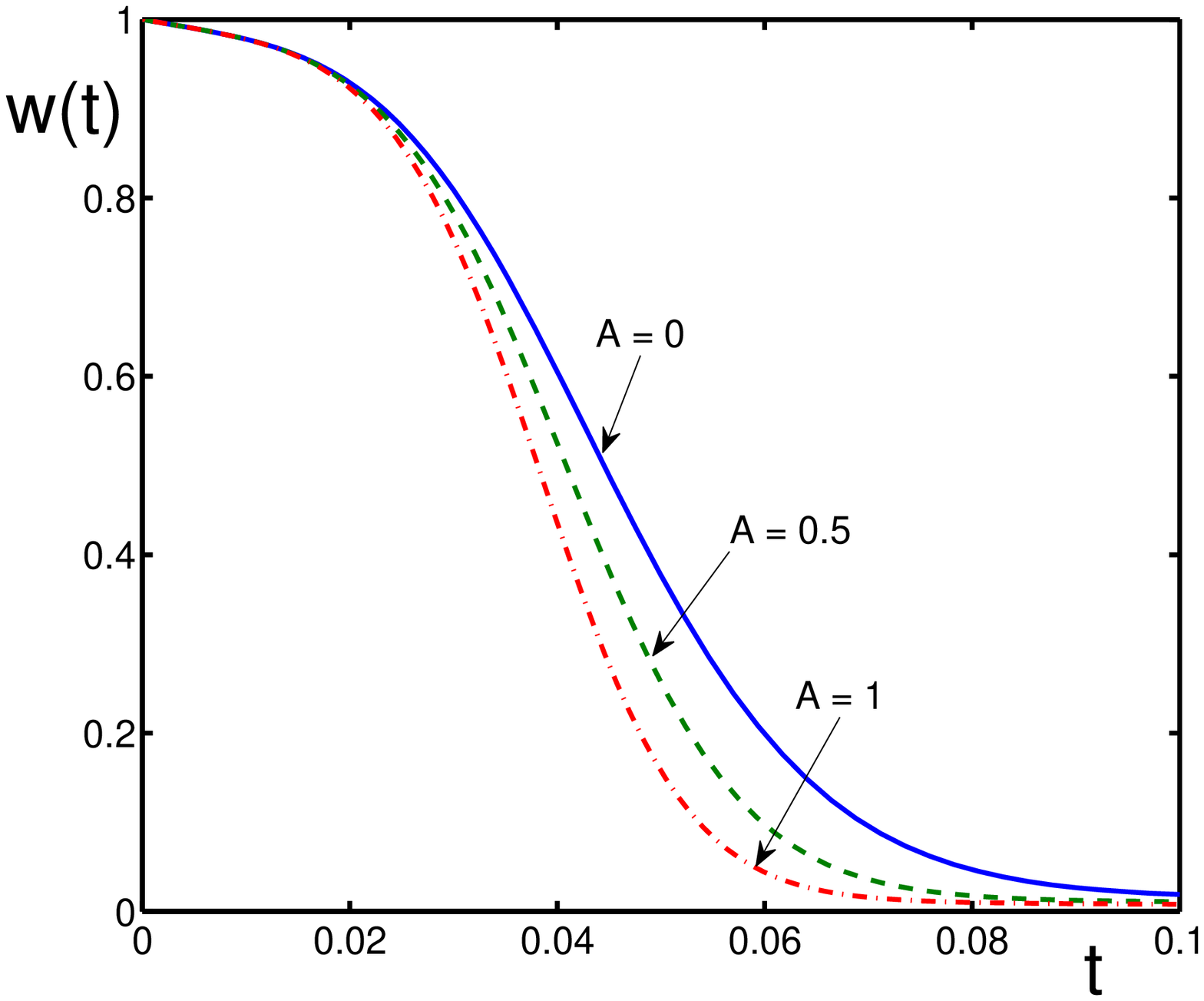} \hspace{1cm}
\includegraphics[width=8cm]{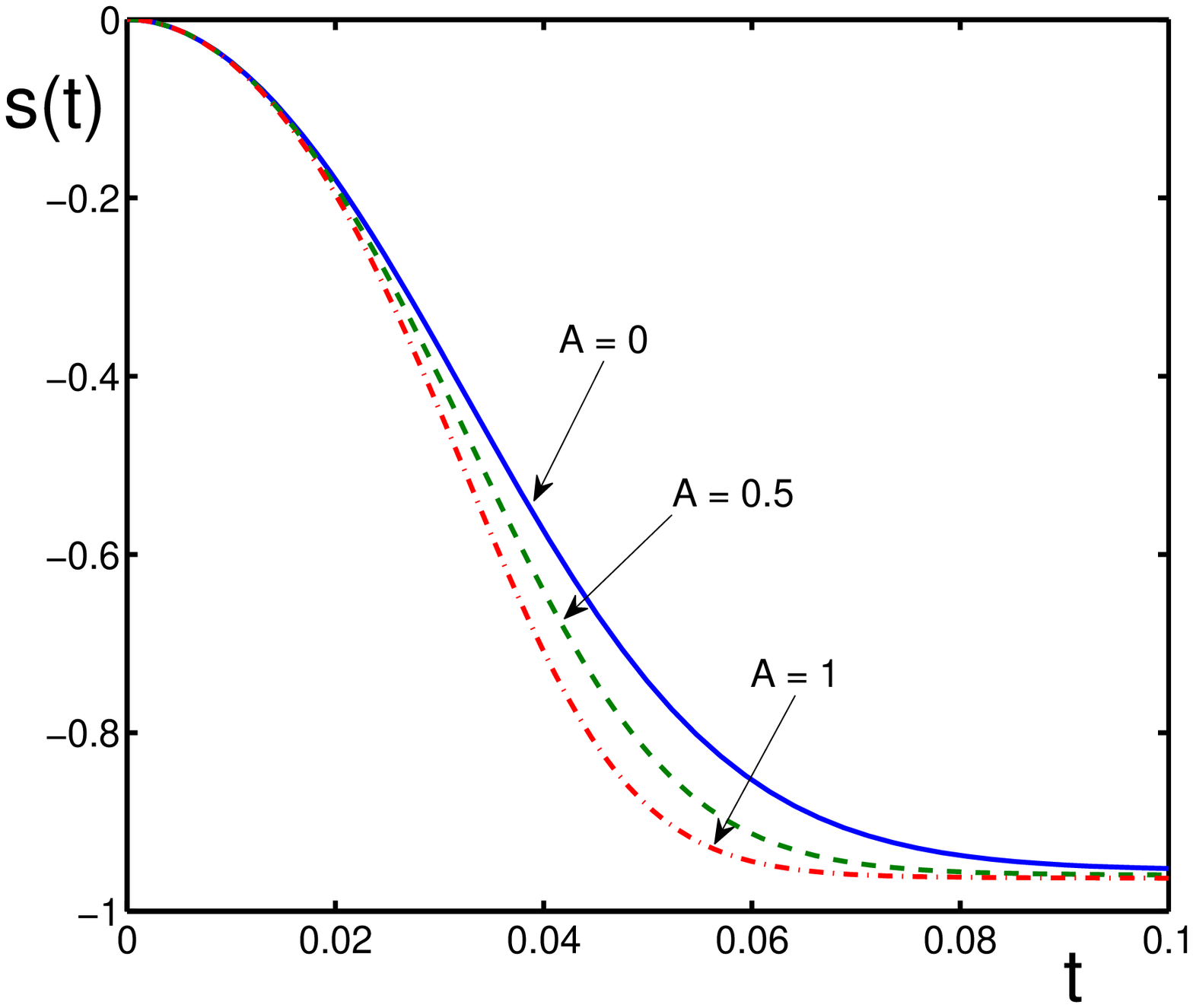} } }
\vspace{9pt}
\centerline{
\hbox{ \includegraphics[width=8cm]{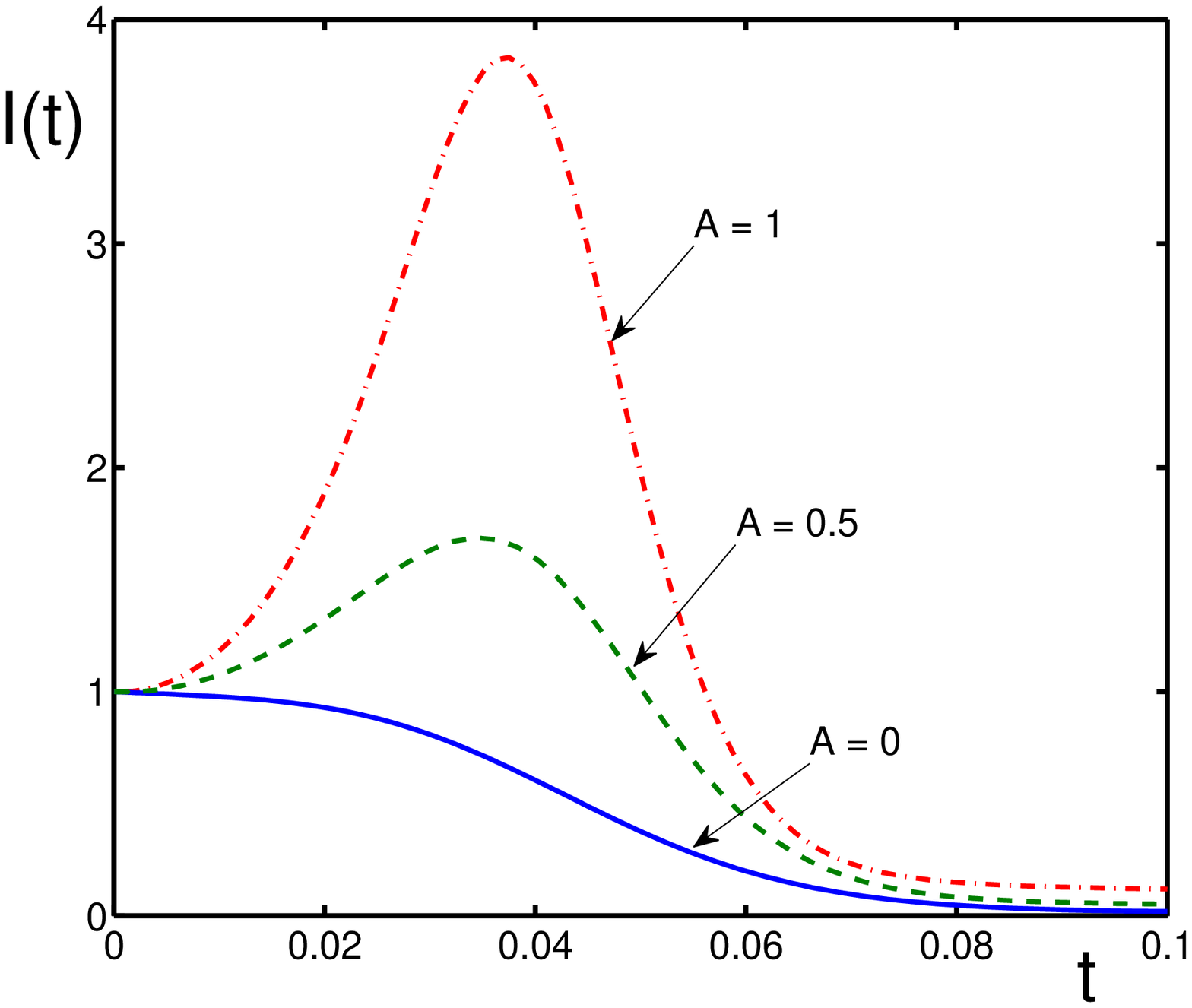} } }
\caption{Regime of spin induction. Coherence intensity $w(t)$, spin 
polarization $s(t)$, and radiation intensity $I(t)$ for the same 
parameters, as in Fig. 1, but for the initial conditions $w_0 = 1$, 
$s_0 = 0$. }
\label{fig:Fig.3}
\end{figure}

\newpage

\begin{figure}[ht]
\vspace{9pt}
\centerline{
\hbox{ \includegraphics[width=8cm]{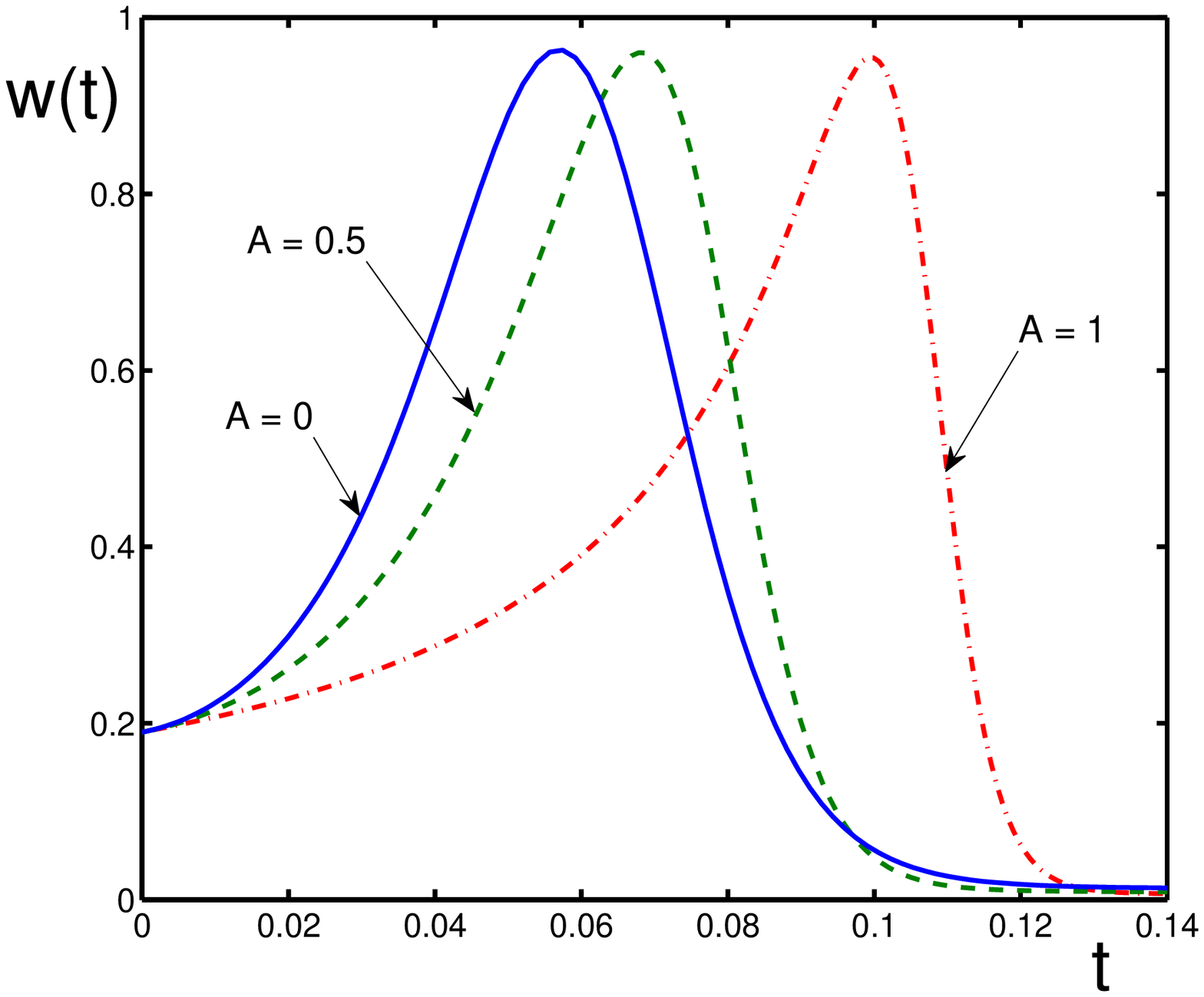} \hspace{1cm}
\includegraphics[width=8cm]{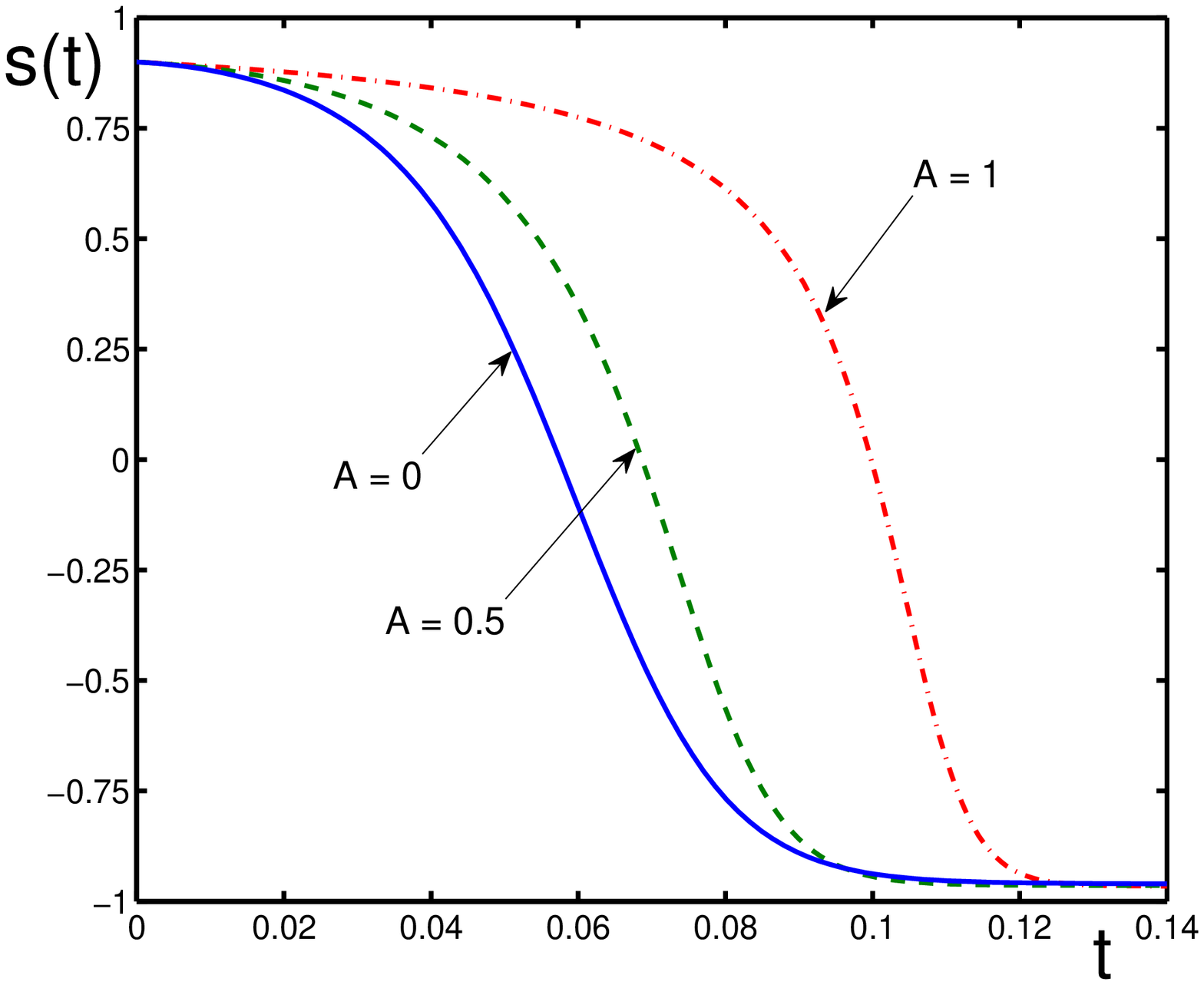} } }
\vspace{9pt}
\centerline{
\hbox{ \includegraphics[width=8cm]{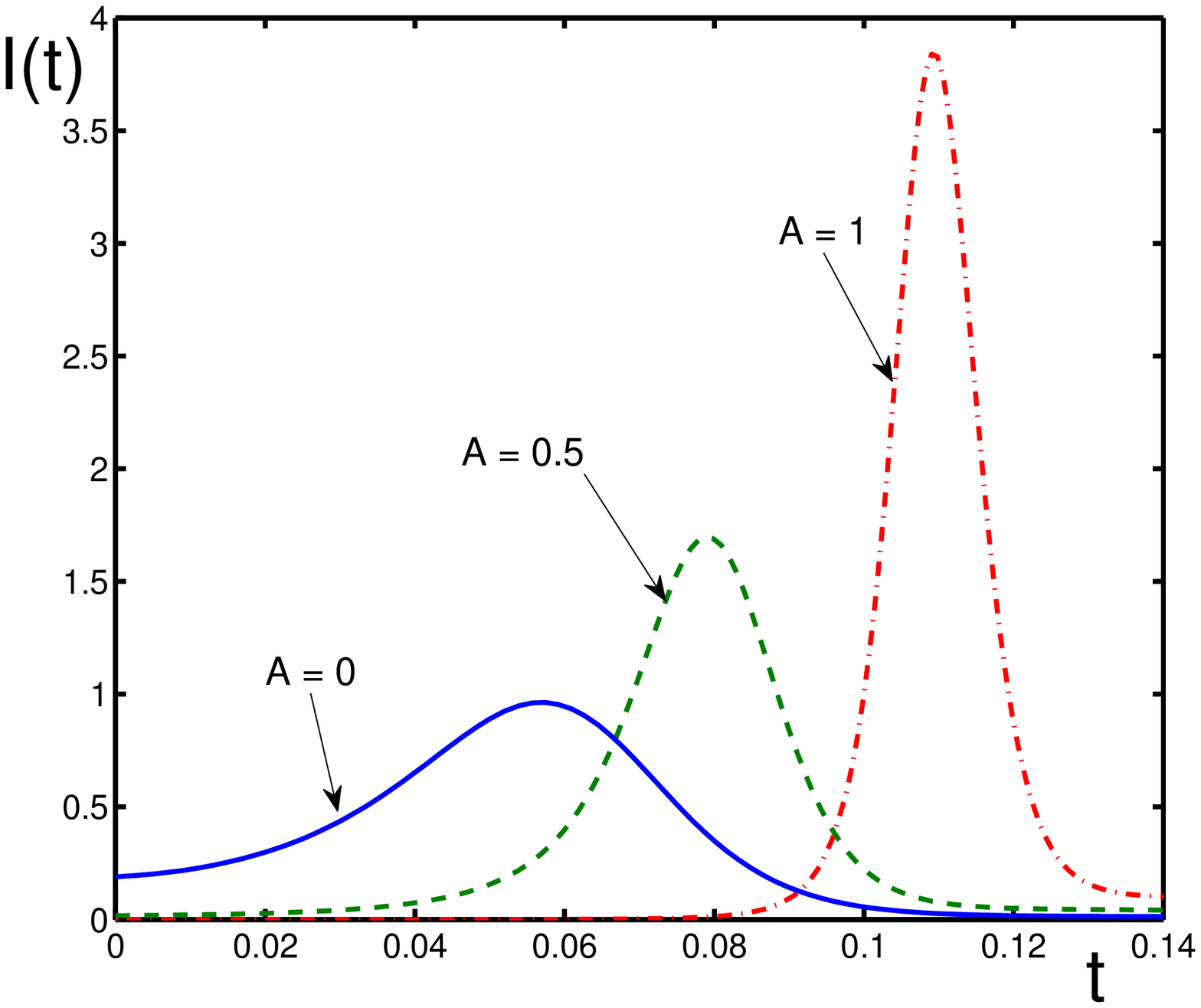} } }
\caption{ Regime of triggered spin superradiance. Coherence intensity 
$w(t)$, spin polarization $s(t)$, and radiation intensity $I(t)$ for 
the same parameters, as in Fig. 1, but for the initial conditions 
$w_0 = 0.19$, $s_0 = 0.9$.}
\label{fig:Fig.4}
\end{figure}

\end{document}